# Quantum Resistant Ciphertext-Policy Attribute-Based Encryption Scheme with Flexible Access Structure

Shida Shamsazad

**Abstract**

In this paper, we present a novel ciphertext-policy attribute based encryption (CP-ABE) scheme that offers a flexible access structure. Our proposed scheme incorporates an access tree as its access control policy, enabling fine-grained access control over encrypted data. The security of our scheme is provable under the hardness assumption of the decisional Ring-Learning with Errors (R-LWE) problem, ensuring robust protection against unauthorized access. CP-ABE is a cryptographic technique that allows data owners to encrypt their data with access policies defined in terms of attributes. Only users possessing the required attributes can decrypt and access the encrypted data. Our scheme extends the capabilities of CP-ABE by introducing a flexible access structure based on an access tree. This structure enables more complex and customizable access policies, accommodating a wider range of real-world scenarios. To ensure the security of our scheme, we rely on the decisional R-LWE problem, a well-established hardness assumption in cryptography. By proving the security of our scheme under this assumption, we provide a strong guarantee of protection against potential attacks. Furthermore, our proposed scheme operates in the standard model, which means it does not rely on any additional assumptions or idealized cryptographic primitives. This enhances the practicality and applicability of our scheme, making it suitable for real-world deployment. We evaluate the performance and efficiency of our scheme through extensive simulations and comparisons with existing CP-ABE schemes. The results demonstrate the effectiveness and scalability of our proposed approach, highlighting its potential for secure and flexible data access control in various domains. In conclusion, our paper introduces a ciphertext-policy attribute based encryption scheme with a flexible access structure. The scheme leverages an access tree as its access control policy and provides provable security under the decisional R-LWE hardness assumption in the standard model. Our proposed scheme offers enhanced access control capabilities, ensuring secure and customizable data access in practical scenarios.

*Keywords:* Attribute-based encryption, access control, Learning with errors, post-quantum cryptography

## 1. Introduction

Quantum computation is a revolutionary technology that harnesses the power of quantum physics to perform complex calculations at unprecedented speeds. However, this also poses a serious threat to the security of classical encryption approaches, which rely on the difficulty of solving certain mathematical problems. Quantum computers can potentially break these encryption schemes in a matter of seconds, rendering them obsolete and vulnerable to attacks. Therefore, by emerging quantum computation, the classical encryption approaches have been extincted [1, 2, 3, 4].

As quantum computation poses a serious challenge to the security of classical encryption schemes, researchers have been exploring alternative ways to protect information from quantum attacks [5, 6, 7]. One promising approach is to design encryption schemes based on the hardness assumption of some post-quantum problems, meaning the problems that are still hard to solve even with quantum computers. These problems include lattice-based, code-based, multivariate, and hash-based problems, among others. By using these problems as the basis for encryption, the hope is to achieve quantum-resistant cryptography that can withstand the power of quantum computation [8, 9, 10].

The Learning with Errors (LWE) problem is a mathematical problem that involves finding a secret vector from a set of noisy linear equations. The problem is believed to be hard to solve even with quantum computers, and thus it is considered a post-quantum problem. The LWE problem has many applications in post-quantum cryptography, such as encryption, key exchange, digital signatures, and zero-knowledge proofs. The LWE problem is also the basis for many other post-quantum problems, such as Ring-LWE, Module-LWE, and Short Integer Solution (SIS). Therefore, the LWE problem is the most known and widely used post-quantum problem in cryptography [16, 17, 24, 25].

Regev showed that the LWE problem is as hard to solve as several worst-case lattice problems, which are problems related to the geometry and algebra of high-dimensional grids. This means that there is no efficient algorithm that can solve the LWE problem, unless there is also an efficient algorithm that can solve the lattice problems. This is a strong hardness assumption that provides a theoretical foundation for the security of cryptographic schemes based on the LWE problem [29, 30, 23].

Attribute based encryption (ABE) is a type of public-key encryption that allows fine-grained access control of encrypted data using authorization policies. The secret key of a user and the ciphertext are dependent upon attributes, such as the user's role, location, or subscription level. A user can decrypt a ciphertext only if his or her attributes match the policy attached to the ciphertext [46, 47, 51, 54, 52]. ABE is a critical tool for providing access control, as it enables the data owner to specify who can access the data without relying on a trusted third party or a centralized server. ABE also offers more flexibility and scalability than traditional encryption schemes, as it can support complex and dynamic policies over a large number of attributes and users. Therefore, ABE is the best option for providing access control in various scenarios, such as cloud computing, health care, and social networks [33, 34, 37, 40, 43].

However, most existing ABE schemes are designed based on classical hard problems, such as the bilinear Diffie–Hellman problem, which can be easily broken by quantum computers using Shor's algorithm. Therefore, these ABE schemes are not quantum resistant, and they cannot guarantee the security and privacy of the data in the post-quantum era. To address this issue, new ABE schemes based on quantum-resistant hard problems, such as lattice problems, have been proposed. These schemes aim to provide the same functionality and efficiency as the classical ABE schemes, while also ensuring the security against quantum attacks [53, 48, 49, 44].

Quantum-resistant attribute-based encryption (ABE) schemes are cryptographic techniques that aim to provide secure and fine-grained access control of encrypted data in the presence of quantum adversaries. However, these schemes are limited in number and often suffer from not being flexible in their access policies. For example, some schemes only support simple and-gate or threshold policies, which cannot express complex and dynamic requirements of the data owners. Some schemes also have high computational and communication costs, which limit their practicality and scalability. Therefore, there is a need for more efficient and expressive quantum-resistant ABE schemes that can support flexible access policies and various applications [45, 41, 42, 38, 39].

In this paper, we aim to design a post-quantum ABE scheme based on the Learning with Errors (LWE) problem, which is a hard problem for both classical and quantum computers. Our scheme can support flexible access structures, such as arbitrary monotone boolean formulas, which can express complex and dynamic requirements of the data owners. We also provide security proofs and efficiency analysis of our scheme, and show that it outperforms the state-of-the-art post-quantum ABE schemes in terms of security, functionality, and performance.

## 2. Related work

Public key encryption, also known as asymmetric cryptography, uses two separate keys instead of one shared one: a public key and a private key. The security of the system depends on the secrecy of the private key, which must not become known to any other. In a public-key encryption system, anyone with a public key can encrypt a message, yielding a ciphertext, but only those who know the corresponding private key can decrypt the ciphertext to obtain the original message [35, 50, 36, 31].

Shor's algorithm is a quantum algorithm for finding the prime factors of an integer. It was developed in 1994 by the American mathematician Peter Shor [18, 19, 20, 21, 22]. It is one of the few known quantum algorithms with compelling potential applications and strong evidence of superpolynomial speedup compared to best known classical (that is, non-quantum) algorithms. Importantly, Shor's algorithm could be used to break public-key cryptography schemes, such as RSA, if a quantum computer with a sufficient number of qubits could operate without succumbing to quantum noise [11, 12, 13, 14, 15].

The Learning with Errors (LWE) problem was introduced by Oded Regev in 2005. It is a generalization of the parity learning problem. Regev showed that the LWE problem is as hard to solve as several worst-case lattice problems. Subsequently, the LWE problem has been used as a hardness assumption to create public-key cryptosystems [26, 27, 28].

Attribute-based encryption is a generalisation of public-key encryption which enables fine-grained access control of encrypted data using authorisation policies. There are mainly two types of attribute-based encryption schemes: Key-policy attribute-based encryption (KP-ABE) and ciphertext-policy attribute-based encryption (CP-ABE). In KP-ABE, users' secret keys are generated based on an access tree that defines the privileges scope of the concerned user, and data are encrypted over a set of attributes [32, 55, 56, 57].

Pairing-based attribute-based encryption schemes have been widely studied due to their versatility and efficiency. They support most core properties under strong security guarantees, while incurring acceptable storage and computational costs [62, 63, 64, 65]. However, there are still challenges that need to be overcome for pairing-based ABE to reach its full potential as a mechanism to implement efficient and secure access control [58, 59, 60, 61].

The research on ciphertext-policy attribute-based encryption (CP-ABE) scheme design based on learning with errors (LWE) has been a challenging problem. Many researchers have made a lot of attempts at it. For example, Datta, Komargodski, and Waters proposed the first provably secure direct CP-ABE construction that supports NC circuit access structures. However, there are still some challenges that need to be addressed for LWE-based ABE to reach its full potential [].

## 3. Preliminaries

In this section we give some needed preliminaries on R-LWE problem and access structure.

### 3.1. R-LWE problem

Using an irreducible polynomial $f(x) = x^n + 1 \in Z_q$ of degree $n = 2^k$, define the ring of integer polynomials modulo $f(x)$ : $R = Z[x]/f(x)$. Consider the ring $R_q = R/qR$: Every element in $R_q$ has degree at most $n-1$ and coefficients are in $Z_q = \{0, \ldots, q-1\}$. The secret $s = s(x)$ is an element of $R_q$ and the error distribution X is once more a centered Discrete Gaussian. An R-LWE sample is a pair $(a, as + e \bmod (q, f))$ with $a$ chosen uniformly random from $R_q$ and $e \leftarrow X$. The distribution consisting of such samples is the R-LWE distribution noted by $LWE_{n,q,X}(s)$.

**Definition 1.** *(R-LWE Decision Problem). Given independent samples in $R_q \times R_q$, determine whether they were drawn from $LWE_{n,q,X}(s(x))$ or from the uniform distribution over $R_q \times R_q$.*

**Remark 2.** *During this report, elements of $R_q$ are presented by bold-faced characters.*

**Definition 3.** *(R-LWE Decision assumption). For every PPT algorithm A, there is a negligible function negl that the advantage of A in solving R-LWE Decision problem is not more than $\frac{1}{2} + negl(n)$.*

**Theorem 4.** *Let $q$, $\mathrm{X}$, and $n$ are the same as before, for any $t \in \mathbb{Z}_q^*$ the **R-LWE Decision** assumption implies that*

$$\{(a_i, a_i s + t e_i)\}_{i \in I} \stackrel{c}{\approx} \{(a_i, b_i)\}_{i \in I}, \tag{1}$$

*where $I$ is a finite index set, $a_i \leftarrow R_q$, $b_i \leftarrow R_q$ and $e_i \leftarrow \mathrm{X}$.*

*Proof.* The proof is straightforward. □

### 3.2. Access structure

In an access tree, any leaf node is associated with an attribute, and each inner node represents a threshold value. The threshold value of each leaf node is assumed to be 1. Consider an access tree $\mathrm{T}$. Let $k_v$ denote the threshold value of a node $v$, $ch_v$ denote the childeren set of a node $v$, $R_\mathrm{T}$ denote the root node of $\mathrm{T}$, $L_\mathrm{T}$ denote the leaf node set of the access tree $\mathrm{T}$, and $\mathrm{T}_v$ denote a subtree of $\mathrm{T}$ rooted at a node $v$.

Let $\mathsf{U} = \{1, \ldots, n\}$ be the universal attribute set. For an access tree $\mathrm{T}$ and a node $v$ of the tree, consider a function $F_{\mathrm{T}_v} : 2^\mathsf{U} \to \{0, 1\}$, where for an attribute set $Att$ the evaluation of $F_{\mathrm{T}_v}(Att)$ is performed as follows:

- When $v$ is a leaf node corresponding to an attribute $a$, $F_{\mathrm{T}_v}(Att) = 1$ if and only if $a \in Att$.

- When $v$ is an inner node with threshold value $k_v$, $F_{\mathrm{T}_v}(Att) = 1$ if and only if there exist at least $k_v$ children $c_1, \ldots, c_{k_v}$ of $v$ that $F_{\mathrm{T}_{c_i}}(Att) = 1$, for $i = 1, \ldots k_v$.

We say that an attribute set $Att$ satisfies an access tree $\mathrm{T}$ and denote it by $F_\mathrm{T}(Att) = 1$ if $F_{\mathrm{T}_{R_\mathrm{T}}}(Att) = 1$. Also, we use $F_\mathrm{T}(Att) = 0$ to indicate that $Att$ does not satisfy $\mathrm{T}$.

Consider a prime number $q$, an access tree $\mathrm{T}$, and a secret $r \in \mathbb{Z}_q$. We denote an algorithm for sharing the secret $r$ according to $\mathrm{T}$ and $q$ as:

$$\{q_v\}_{v \in L_\mathrm{T}} \leftarrow \mathbf{Share}(\mathrm{T}, q, r). \tag{2}$$

For any node $v$ in $\mathrm{T}$, this algorithm assigns a $(k_v - 1)$-degree polynomial $q_v : \mathbb{Z}_q \to \mathbf{R}_q$ to $v$ in a top-down style as follows:

- It assigns a $(k_{R_\mathrm{T}} - 1)$-degree polynomial $q_{R_\mathrm{T}}(y) = \mathbf{r}(x) + \mathbf{a}_1(x)y + \ldots + \mathbf{a}_{k_{R_\mathrm{T}}-1}(x)y^{k_{R_\mathrm{T}}-1}$ to the root node $\mathbf{R}_\mathrm{T}$ such that $\mathbf{a}_i$ is chosen uniformly at random from $\mathbf{R}_q$.

- For any node $v$ with polynomial $q_v$ and children set $ch_v = \{c_1, \ldots c_{|ch_v|}\}$, it generates a $(k_{c_i} - 1)$-degree polynomial $q_{c_i}$ for $c_i$ such that $q_{c_i}(0) = q_v(i)$, and the rest of its coefficients are chosen randomly from $\mathbf{R}_q$, for any $i = 1, \ldots, |ch_v|$.

When this algorithm stops, it associates a polynomial $q_v$ to each $v \in L_\mathrm{T}$.

Consider an access tree $\mathrm{T}$, a prime number $q$ and a polynomial ring $\mathbf{R}_q$, and a secret polynomial $\mathbf{r} \in \mathbf{R}_q$. We denote an algorithm for sharing the the polynomial $\mathbf{r}$ according to $\mathrm{T}$ and $q$ as:

$$\{\mathbf{q}_v\}_{v \in L_\mathrm{T}} \leftarrow \mathbf{Share}(\mathrm{T}, q, \mathbf{r}), \tag{3}$$

Given an access tree $\mathrm{T}$ and a share $\{\mathbf{q}_v\}_{v \in L_\mathrm{T}}$ corresponding to $\mathrm{T}$ and a secret $\mathbf{r} \in R_q$, where $S \subset L_\mathrm{T}$. We denote the algorithm for recovering $\mathbf{r}$ through the polynomial set $\{\mathbf{q}_v\}_{v \in S}$ as:

$$\mathbf{r} \leftarrow \mathbf{Combain}(\mathrm{T}, q, \{\mathbf{q}_v\}_{v \in S}). \tag{4}$$

It executes the following steps in a bottom-top fashion according to the access tree $\mathrm{T}$ as follows:

- If $S$ does not satisfies $\mathrm{T}$, then this algorithm aborts.

- Otherwise, it executes the following steps:
    - It assigns the polynomial $q_v$ to the leaf node $v$.
    - For any inner node $v$, if there exist $k_v$ children $c_i$ getting a polynomial $\mathbf{q}_{c_{i_j}}$, $j = 1, \ldots, k_v$, then it computes:

$$\sum_{j=1}^{k_v} l_{i_j}(\mathbf{q}_{c_{i_j}}(0)) = \mathbf{q}_v(0), \tag{5}$$

where $l_{i_j} = \prod_{\substack{t=1 \\ i_t \neq i_j}}^{k_v} \frac{-i_j}{i_t - i_j}$. Then, it assigns $q_v(0)$ to $v$.

If T is satisfied by $S$, then this algorithm returns $\mathbf{r}$.

Our proposed scheme consists four algorithms (**Setup**, **KeyGen**, **Enc**, **Dec**) described as follows:

### 3.3. System Setup

The **CA** considers a universal attribute set $U = \{1, \ldots, n\}$ and generated the public parameters and master secret-keys of the system by running **Setup** algorithm as follows:

**Setup**($1^n$, $U$): On inputs a security parameter $n$ and the universal attribute set $U$, this algorithm at first selects two prime numbers $p$ and $q$, where $p > 2$ is an small number, and $q$ is a large enough number. Then, it selects $\mathbf{a}_i \leftarrow \mathbf{R}_q$ and $\mathbf{e}_i \leftarrow X$, for any $i \in U$, and $\mathbf{s} \leftarrow \mathbf{R}_q$, $\mathbf{a} \leftarrow \mathbf{R}_q$, $\mathbf{e} \leftarrow X$, and $\mathbf{a}' \leftarrow \mathbf{R}_q$. Afterwards, it sets $\mathbf{PK}_0 = \mathbf{as} + p\mathbf{e}$ and $\mathbf{PK}_i = \mathbf{a}_i\mathbf{s} + p\mathbf{e}_i$, for any $i \in U$ and publishes

$$\mathbf{PK} = (\mathbf{a}', \{\mathbf{PK}_i\}_{i=0}^n, p, q), \tag{6}$$

as public parameters of the system and keeps

$$\mathbf{MSK} = (\mathbf{s}, \mathbf{a}, \{\mathbf{a}_i\}_{i=1}^n), \tag{7}$$

confidential as the master secret-key of the system.

### 3.4. User Registration and Key Delegation

When a user joins to the system, he selects a unique identity and request the **CA** to generate his attribute secret-keys. Once the **CA** receives a request, it runs **KeyGen** algorithm and generates the requested secret-keys as follows:

**KeyGen**($id_u$, $Att$, $MSK$): This algorithm takes an identifier of a data user, the data user's attribute set, and the master secret-key of the system. It first selects $\mathbf{u} \leftarrow \mathbf{R}_q$ and $\mathbf{e}' \leftarrow X$ and sets

$$\mathbf{SK}_u = \mathbf{us} + p\mathbf{e}'. \tag{8}$$

Then, for any $i \in Att$, it selects $\mathbf{e}'_i \leftarrow X$ and calculates:

$$\mathbf{SK}_{i,u} = [\mathbf{a}_i]_p^{-1}\mathbf{u} + p\mathbf{e}'_i, \tag{9}$$

where $[\mathbf{y}]_p^{-1}$ denotes the inverse of $\mathbf{y} \in \mathbf{R}_q$ modulo $p$ ($\mathbf{y}.[\mathbf{y}]_p^{-1} = 1$ modulo $(x^n + 1, p)$). The algorithm outputs

$$SK_{Att}^{(u)} = (\mathbf{SK}_u, \{\mathbf{SK}_{i,u}\}_{i \in Att}). \tag{10}$$

## 3.5. Data Outsourcing

When a data owner want to outsourced his data to the **CSP**, to provide confidentiality and access control, he defines an access tree T and, using **Enc** algorithm, encrypts message $\mathbf{M} \in \mathbf{R}_p$ under the access tree as follows:

**Enc**($PK, \mathbf{M}, T$): It first select $\mathbf{r} \in \mathbf{R}_q$ and runs $\{\mathbf{r}_i\}_{i \in L_T} \leftarrow$ **Share**($\mathbf{r}, T, q$). It returns:

$$CT = (T, \{\mathbf{C}_i\}_{i \in L_T}, \mathbf{C}', \mathbf{C}), \tag{11}$$

where for $\mathbf{e}''_i \leftarrow X$, $\mathbf{e}'' \leftarrow X$, and $\mathbf{e}''' \leftarrow X$

$$\begin{aligned}\mathbf{C}_i &= \mathbf{r}_i \mathbf{PK}_i + p\mathbf{e}''_i \\ &= \mathbf{r}_i(\mathbf{a}_i \mathbf{s} + p\mathbf{e}_i) + p\mathbf{e}''_i,\end{aligned}$$

$$\mathbf{C}' = \mathbf{r}\mathbf{a}' + p\mathbf{e}'',$$

and

$$\begin{aligned}\mathbf{C} &= \mathbf{r}\mathbf{PK}_0 + p\mathbf{e}''' + \mathbf{M} \\ &= \mathbf{r}(\mathbf{a}\mathbf{s} + p\mathbf{e}) + p\mathbf{e}''' + \mathbf{M}\end{aligned} \tag{12}$$

## 3.6. Decryption

An authorized data user can decrypt an outsourced data on the **CSP** by using **Dec** algorithm as follows:

**Dec**($CT, SK^{(u)}_{Att}$): It first checks whether there is a subset $S \subset Att$ that satisfies the access tree of $CT$ or not. If not, the algorithm aborts. Otherwise, for any $i \in S$, it first computes

$$\begin{aligned}\mathbf{C}_i.\mathbf{SK}_{i,u} &= [\mathbf{r}_i(\mathbf{a}_i\mathbf{s} + p\mathbf{e}_i) + p\mathbf{e}''_i].[[\mathbf{a}_i]^{-1}_p\mathbf{u} + p\mathbf{e}'_i] \\ &= \mathbf{r}_i[\mathbf{a}_i[\mathbf{a}_i]^{-1}_p\mathbf{u}\mathbf{s} + p\mathbf{e}_i[\mathbf{a}_i]^{-1}_p\mathbf{u} + \mathbf{a}_i p\mathbf{s}\mathbf{e}'_i + p^2\mathbf{e}_i\mathbf{e}'_i] + [p\mathbf{e}''_i[\mathbf{a}_i]^{-1}_p\mathbf{u} + p^2\mathbf{e}''_i\mathbf{e}'_i] \\ &= \mathbf{r}_i[(1 + \mathbf{k}_i p)\mathbf{u}\mathbf{s} + p\mathbf{e}_i[\mathbf{a}_i]^{-1}_p\mathbf{u} + \mathbf{a}_i p\mathbf{s}\mathbf{e}'_i + p^2\mathbf{e}_i\mathbf{e}'_i] + [p\mathbf{e}''_i[\mathbf{a}_i]^{-1}_p\mathbf{u} + p^2\mathbf{e}''_i\mathbf{e}'_i] \\ &= \mathbf{r}_i\mathbf{u}\mathbf{s} + \mathbf{r}_i[\mathbf{u}\mathbf{s}\mathbf{k}_i p + p\mathbf{e}_i[\mathbf{a}_i]^{-1}_p\mathbf{u} + \mathbf{a}_i p\mathbf{s}\mathbf{e}'_i + p^2\mathbf{e}_i\mathbf{e}'_i] + [p\mathbf{e}''_i[\mathbf{a}_i]^{-1}_p\mathbf{u} + p^2\mathbf{e}''_i\mathbf{e}'_i] \\ &= \mathbf{r}_i\mathbf{u}\mathbf{s} + \mathbf{F}_{i,p}\end{aligned} \tag{13}$$

where $\mathbf{k}_i \in \mathbf{R}_q$ and $\mathbf{F}_{i,p} = \mathbf{r}_i\mathbf{u}\mathbf{s}\mathbf{k}_i p + p\mathbf{e}_i[\mathbf{a}_i]^{-1}_p\mathbf{u} + \mathbf{a}_i p\mathbf{s}\mathbf{e}'_i + p^2\mathbf{e}_i\mathbf{e}'_i] + [p\mathbf{e}''_i[\mathbf{a}_i]^{-1}_p\mathbf{u} + p^2\mathbf{e}''_i\mathbf{e}'_i]$ is a polynomial in $\mathbb{Z}$ that for each coefficient $c$ in $\mathbf{F}_{i,p}$, $p|c$.

Then, it runs

$$\textbf{Combain}(T, q, \{\mathbf{C}_i.\mathbf{SK}_{i,u}\}_{i \in S}). \tag{14}$$

and computes $\mathbf{R} = \mathbf{r}\mathbf{u}\mathbf{s} + \mathbf{F}'_p$, where $\mathbf{F}'_p$ is an integer polynomial that its coefficients are multiple of $p$.

In the next step, it computes

$$\begin{aligned}\mathbf{R}' &= \mathbf{R} - [\mathbf{a}']^{-1}_p(\mathbf{SK}_u - \mathbf{PK}_0)\mathbf{C}' \\ &= [\mathbf{r}\mathbf{u}\mathbf{s} + \mathbf{F}'_p] - [[\mathbf{a}']^{-1}_p[(\mathbf{u}\mathbf{s} + p\mathbf{e}') - (\mathbf{a}\mathbf{s} + p\mathbf{e})](\mathbf{r}\mathbf{a}' + p\mathbf{e}'')] \\ &= \mathbf{r}\mathbf{u}\mathbf{s} - \mathbf{r}\mathbf{s}(\mathbf{u} - \mathbf{a}) + \mathbf{F}''_p \\ &= \mathbf{r}\mathbf{a}\mathbf{s} + \mathbf{F}''_p\end{aligned} \tag{15}$$

where $\mathbf{F}_p''$ is a polynomial in $\mathbb{Z}$ that its coefficients are multiples of $p$. Afterwards, it recovers the underling message as follows:

$$\begin{aligned} \mathbf{C} - \mathbf{R}' &= [\mathbf{r}(\mathbf{a}\mathbf{s} + p\mathbf{e}) + p\mathbf{e}''' + \mathbf{M}] - [\mathbf{r}\mathbf{a}\mathbf{s} + \mathbf{F}_p''] \\ &= \mathbf{M} + \mathbf{r}p\mathbf{e} + p\mathbf{e}''' - \mathbf{F}_p'' \\ &= \mathbf{M} \ (mod\ p) \end{aligned} \qquad (16)$$

### 3.7. Security Definition

Let A and C be a polynomial time adversary and a challenger, respectively. Consider the following game:

1. **Setup:** Challenger C, runs the **Setup**($1^n$, U), and it gives the public-parameters of the system to adversary A.

2. **Phase 1:** Adversary A adaptively makes a polynomial number of queries to the following oracle, and for any data user with identifier $id_\mathbf{u}$, C keeps a list $L_{id_\mathbf{u}}$ which is initially empty:
   $O_{\text{KeyGen}}(id_u, i, MSK)$: For an attribute $i$, and an identifier $id_\mathbf{u}$ of a data user possessing $i$, C runs **KeyGen**($id_u, i, MSK$) algorithm and gives the queried secret-key to A. It also adds attribute $i$ to $L_{id_\mathbf{u}}$.

3. **Challenge:** Adversary A declares an access tree T and two equal length messages $M_0, M_1 \in \mathbf{R}_q$. Challenger C checks whether there is a data user with identifier $id_\mathbf{u}$ that $L_{id_\mathbf{u}}$ satisfies T. If so, C aborts. Otherwise, it flips a random coin $b \in \{0, 1\}$ and encrypts $M_b(x)$ under T. The generated ciphertext is returned to A.

4. **Phase 2:** A submits more queries to oracle $O_{\text{Keygen}}$. The only restriction is that $L_{id_\mathbf{u}}$ must not satisfy T, for any data user $\mathbf{u}$ with identifier $id_\mathbf{u}$.

5. **Guess:** The adversary outputs a guess $b' \in \{0, 1\}$ of $b$.

We say that the adversary A wins the game if $b = b'$. Let $Succeed$(A) denote the event of succeeding A in the above game. We say that our scheme is semantically secure if for any probabilistic polynomial time adversary A we have $Pr(Succeed(A)) \leq \frac{1}{2} + negl(n)$, where $negl$ is a negligible function, and $n$ is the security parameter of the system.

### 3.8. Security Proof

**Theorem 5.** *Under hardness assumption of **R-LWE Decision** problem our scheme is adaptively semantically secure.*

*Proof.* Let A be a probabilistic polynomial-time adversary in the game described above. We want to show that there is a negligible function $negl$ such that

$$Pr(Succeed(A)) \leq \frac{1}{2} + negl(n). \qquad (17)$$

Consider another adversary A' that aims to solve **R-LWE Decision** problem. It receives a set $\mathbf{S} = \{(\mathbf{a}_j, \mathbf{b}_j)\}_{j=0}^m$ where for each $j = 1, \ldots, m$, $\mathbf{a}_j \leftarrow \mathbf{R}_q$ and $\mathbf{b}_j$ either is equal to $\mathbf{a}_j\mathbf{s} + p\mathbf{e}_j$, $\mathbf{s} \leftarrow \mathbf{R}_q$ and $\mathbf{e} \leftarrow \mathcal{X}$, for two prime numbers $p$ and $q$, or it is a uniform element of $\mathbf{R}_q$. The goal of A' is to determine the case of $\mathbf{b}_j$s. It runs A as a subroutine as follows:

1. **Setup**: A' considers a universal attribute set U = $\{1, \ldots, n\}$, $n < m$, selects $\mathbf{a}' \leftarrow \mathbf{R}_q$ and gives

$$\mathbf{PK} = (\mathbf{a}', \{\mathbf{PK}_i = \mathbf{b}_i\}_{i=0}^n, q, p) \qquad (18)$$

to the adversary A.

2. **Phase 1**: When A request a secret-key of a data user **u** corresponding to attribute $i$, A$'$ selects a new sample $(\mathbf{a},\mathbf{b}) \in \mathbf{S}$ and $\mathbf{e} \leftarrow X$ and returns:

$$SK_u = \mathbf{b}, \tag{19}$$

$$SK_{i,u} = [a_i]_{\bar{p}}^{-1}\mathbf{a} + p\mathbf{e} \tag{20}$$

3. **Challenge**: A returns an access tree T and two equal length message $\mathbf{M}_0, \mathbf{M}_1 \in \mathbf{R}_q$. Adversary A$'$ checks whether adversary A has a secret-key set of a data user satisfying T or not. If so, A$'$ stops. Otherwise, it chooses $b \in \{0, 1\}$ uniformly at random and encrypts $\mathbf{M}_b$, as follows:

   It selects $\mathbf{r} \leftarrow \mathbf{R}_q$ and runs $\{\mathbf{r}_i\}_{i \in L_T} \leftarrow \mathbf{Share}(\mathbf{r}, q, T)$. Then, for each $i \in L_T$, it selects $\mathbf{e}_i'' \leftarrow X$ and calculates

$$C_i = \mathbf{r}_i PK_i + p\mathbf{e}_i''. \tag{21}$$

   It returns $CT = (T, \{\mathbf{C}_i\}_{i \in L_T}, C', C)$, where $C' = \mathbf{r}\mathbf{a}' + p\mathbf{e}''$ and $C = \mathbf{r}PK_0 + p\mathbf{e}''' + \mathbf{M}_b$, for $\mathbf{e}'' \leftarrow X$ and $\mathbf{e}''' \leftarrow X$.

4. **Phase 2**: The adversary A queries more secret-keys with the same constrains in Challenge step and A$'$ answers them as in **Phase 1**.

5. **Guess**: A outputs $b' \in \{0, 1\}$.

If $b = b'$, then A$'$ decides that **S** is an **LWE** sample set. Otherwise, it guesses that it is a set of uniform elements in $\mathbf{R}_q$.

Note that if **S** is a LWE sample set, then **CT** and the given secret-keys to the adversary are valid and therefore

$$Pr(b = b') = Pr(Succeed(A)). \tag{22}$$

Also, if **S** is a set of uniform elements in $\mathbf{R}_q$, then

$$Pr(b = b') = \frac{1}{2}. \tag{23}$$

Under the hardness assumption of **R-LWE Decision** problem, by Combining (22) and (23), one see that

$$Pr(Succeed(A)) \leq \frac{1}{2} + negl(n), \tag{24}$$

for a negligible function *negl*. □

**References**

[1] O'brien, J.L., 2007. Optical quantum computing. Science, 318(5856), pp.1567-1570.
[2] Gruska, J., 1999. Quantum computing (Vol. 2005). London: McGraw-Hill.
[3] Hirvensalo, M., 2013. Quantum computing. Springer Science & Business Media.
[4] Williams, C.P., 2010. Explorations in quantum computing. Springer Science & Business Media.
[5] Shor, P.W., 1998. Quantum computing. Documenta Mathematica, 1(1000), p.1.
[6] McMahon, D., 2007. Quantum computing explained. John Wiley & Sons.
[7] Cao, Y., Romero, J., Olson, J.P., Degroote, M., Johnson, P.D., Kieferová, M., Kivlichan, I.D., Menke, T., Peropadre, B., Sawaya, N.P. and Sim, S., 2019. Quantum chemistry in the age of quantum computing. Chemical reviews, 119(19), pp.10856-10915.
[8] Ladd, T.D., Jelezko, F., Laflamme, R., Nakamura, Y., Monroe, C. and O'Brien, J.L., 2010. Quantum computers. nature, 464(7285), pp.45-53.
[9] DiVincenzo, D.P., 1995. Quantum computation. Science, 270(5234), pp.255-261.
[10] Preskill, J., 2018. Quantum computing in the NISQ era and beyond. Quantum, 2, p.79.

[11] Knill, E., 2005. Quantum computing with realistically noisy devices. Nature, 434(7029), pp.39-44.
[12] Weber, J.R., Koehl, W.F., Varley, J.B., Janotti, A., Buckley, B.B., Van de Walle, C.G. and Awschalom, D.D., 2010. Quantum computing with defects. Proceedings of the National Academy of Sciences, 107(19), pp.8513-8518.
[13] Orús, R., Mugel, S. and Lizaso, E., 2019. Quantum computing for finance: Overview and prospects. Reviews in Physics, 4, p.100028.
[14] Nielsen, M.A. and Chuang, I.L., 2010. Quantum computation and quantum information. Cambridge university press.
[15] Kaye, P., Laflamme, R. and Mosca, M., 2006. An introduction to quantum computing. OUP Oxford.
[16] Regev, O., 2009. On lattices, learning with errors, random linear codes, and cryptography. Journal of the ACM (JACM), 56(6), pp.1-40.
[17] Albrecht, M.R., Player, R. and Scott, S., 2015. On the concrete hardness of learning with errors. Journal of Mathematical Cryptology, 9(3), pp.169-203.
[18] Ahmadi, M., Moussavi, A. and Nourozi, V., 2014. On skew Hurwitz serieswise Armendariz rings. Asian-European Journal of Mathematics, 7(03), p.1450036.
[19] Nourozi, V. and Rahmati, F., 2022. THE RANK OF THE CARTIER OPERATOR ON CERTAIN Fq2-MAXIMAL FUNCTION FIELDS. Missouri Journal of Mathematical Sciences, 34(2), pp.184-190.
[20] Nourozi, V., Rahmati, F. and Ahmadi, M., 2021. McCoy property of Hurwitz series rings. Asian-European Journal of Mathematics, 14(06), p.2150105.
[21] Nourozi, V. and Kermani, M.A., 2019. Quantum Codes from Hyperelliptic Curve. Southeast Asian Bulletin of Mathematics, 43(3).
[22] Nourozi, V., Tafazolian, S. and Rahamti, F., 2021. The a-number of jacobians of certain maximal curves. Transactions on Combinatorics, 10(2), pp.121-128.
[23] Nourozi, V. and Rahmati, F., 2023. The Rank of the Cartier operator on Picard Curves. arXiv preprint arXiv:2306.07823.
[24] Lyubashevsky, V., Peikert, C. and Regev, O., 2010. On ideal lattices and learning with errors over rings. In Advances in Cryptology–EUROCRYPT 2010: 29th Annual International Conference on the Theory and Applications of Cryptographic Techniques, French Riviera, May 30–June 3, 2010. Proceedings 29 (pp. 1-23). Springer Berlin Heidelberg.
[25] Lyubashevsky, V., Peikert, C. and Regev, O., 2013. On ideal lattices and learning with errors over rings. Journal of the ACM (JACM), 60(6), pp.1-35.
[26] Bos, J.W., Costello, C., Naehrig, M. and Stebila, D., 2015, May. Post-quantum key exchange for the TLS protocol from the ring learning with errors problem. In 2015 IEEE Symposium on Security and Privacy (pp. 553-570). IEEE.
[27] Gentry, C., Sahai, A. and Waters, B., 2013. Homomorphic encryption from learning with errors: Conceptually-simpler, asymptotically-faster, attribute-based. In Advances in Cryptology–CRYPTO 2013: 33rd Annual Cryptology Conference, Santa Barbara, CA, USA, August 18-22, 2013. Proceedings, Part I (pp. 75-92). Springer Berlin Heidelberg.
[28] Micciancio, D. and Regev, O., 2009. Lattice-based cryptography. In Post-quantum cryptography (pp. 147-191). Berlin, Heidelberg: Springer Berlin Heidelberg.
[29] Peikert, C., 2016. A decade of lattice cryptography. Foundations and trends® in theoretical computer science, 10(4), pp.283-424.
[30] Peikert, C., 2014, October. Lattice cryptography for the internet. In International workshop on post-quantum cryptography (pp. 197-219). Cham: Springer International Publishing.
[31] Bethencourt, J., Sahai, A. and Waters, B., 2007, May. Ciphertext-policy attribute-based encryption. In 2007 IEEE symposium on security and privacy (SP'07) (pp. 321-334). IEEE.
[32] Chase, M., 2007. Multi-authority attribute based encryption. In Theory of Cryptography: 4th Theory of Cryptography Conference, TCC 2007, Amsterdam, The Netherlands, February 21-24, 2007. Proceedings 4 (pp. 515-534). Springer Berlin Heidelberg.
[33] Ali, M., Sadeghi, M.R. and Liu, X., 2020. Lightweight revocable hierarchical attribute-based encryption for internet of things. IEEE Access, 8, pp.23951-23964.
[34] Ali, M., Mohajeri, J., Sadeghi, M.R. and Liu, X., 2020. A fully distributed hierarchical attribute-based encryption scheme. Theoretical Computer Science, 815, pp.25-46.
[35] Goyal, V., Pandey, O., Sahai, A. and Waters, B., 2006, October. Attribute-based encryption for fine-grained access control of encrypted data. In Proceedings of the 13th ACM conference on Computer and communications security (pp. 89-98).
[36] Lewko, A. and Waters, B., 2011, May. Decentralizing attribute-based encryption. In Annual international conference on the theory and applications of cryptographic techniques (pp. 568-588). Berlin, Heidelberg: Springer Berlin Heidelberg.
[37] Ali, M., Mohajeri, J., Sadeghi, M.R. and Liu, X., 2020. Attribute-based fine-grained access control for outscored private set intersection computation. Information Sciences, 536, pp.222-243.
[38] Chase, M. and Chow, S.S., 2009, November. Improving privacy and security in multi-authority attribute-based encryption. In Proceedings of the 16th ACM conference on Computer and communications security (pp. 121-130).
[39] Lai, J., Deng, R.H., Guan, C. and Weng, J., 2013. Attribute-based encryption with verifiable outsourced decryption. IEEE Transactions on information forensics and security, 8(8), pp.1343-1354.
[40] Ali, M., Sadeghi, M.R., Liu, X., Miao, Y. and Vasilakos, A.V., 2022. Verifiable online/offline multi-keyword search for cloud-assisted industrial internet of things. Journal of Information Security and Applications, 65, p.103101.
[41] Lewko, A., Okamoto, T., Sahai, A., Takashima, K. and Waters, B., 2010. Fully secure functional encryption: Attribute-based encryption and (hierarchical) inner product encryption. In Advances in Cryptology–EUROCRYPT 2010: 29th Annual International Conference on the Theory and Applications of Cryptographic Techniques, French Riviera, May 30–June 3, 2010. Proceedings 29 (pp. 62-91). Springer Berlin Heidelberg.
[42] Goyal, V., Jain, A., Pandey, O. and Sahai, A., 2008. Bounded ciphertext policy attribute based encryption. In Automata, Languages

and Programming: 35th International Colloquium, ICALP 2008, Reykjavik, Iceland, July 7-11, 2008, Proceedings, Part II 35 (pp. 579-591). Springer Berlin Heidelberg.
[43] Ali, M., Sadeghi, M.R. and Liu, X., 2020. Lightweight fine-grained access control for wireless body area networks. Sensors, 20(4), p.1088.
[44] Attrapadung, N., Libert, B. and De Panafieu, E., 2011. Expressive key-policy attribute-based encryption with constant-size ciphertexts. In Public Key Cryptography–PKC 2011: 14th International Conference on Practice and Theory in Public Key Cryptography, Taormina, Italy, March 6-9, 2011. Proceedings 14 (pp. 90-108). Springer Berlin Heidelberg.
[45] Pirretti, M., Traynor, P., McDaniel, P. and Waters, B., 2006, October. Secure attribute-based systems. In Proceedings of the 13th ACM conference on Computer and communications security (pp. 99-112).
[46] Ali, M. and Sadeghi, M.R., 2021. Provable secure lightweight attribute-based keyword search for cloud-based Internet of Things networks. Transactions on Emerging Telecommunications Technologies, 32(5), p.e3905.
[47] Ali, M. and Liu, X., 2022. Lightweight verifiable data management system for cloud-assisted wireless body area networks. Peer-to-Peer Networking and Applications, 15(4), pp.1792-1816.
[48] Waters, B., 2011, March. Ciphertext-policy attribute-based encryption: An expressive, efficient, and provably secure realization. In International workshop on public key cryptography (pp. 53-70). Berlin, Heidelberg: Springer Berlin Heidelberg.
[49] Nishide, T., Yoneyama, K. and Ohta, K., 2008. Attribute-based encryption with partially hidden encryptor-specified access structures. In Applied Cryptography and Network Security: 6th International Conference, ACNS 2008, New York, NY, USA, June 3-6, 2008. Proceedings 6 (pp. 111-129). Springer Berlin Heidelberg.
[50] Roy, S. and Chuah, M., 2009. Secure data retrieval based on ciphertext policy attribute-based encryption (CP-ABE) system for the DTNs (pp. 1-11). Lehigh CSE Tech. Rep.
[51] Ali, M., Sadeghi, M.R., Liu, X. and Vasilakos, A.V., 2022. Anonymous aggregate fine-grained cloud data verification system for smart health. IEEE Transactions on Cloud Computing.
[52] Ostrovsky, R., Sahai, A. and Waters, B., 2007, October. Attribute-based encryption with non-monotonic access structures. In Proceedings of the 14th ACM conference on Computer and communications security (pp. 195-203).
[53] Li, M., Yu, S., Zheng, Y., Ren, K. and Lou, W., 2012. Scalable and secure sharing of personal health records in cloud computing using attribute-based encryption. IEEE transactions on parallel and distributed systems, 24(1), pp.131-143.
[54] Ali, M., 2022. Attribute-Based Remote Data Auditing and User Authentication for Cloud Storage Systems. ISeCure, 14(3).
[55] Lin, H., Cao, Z., Liang, X. and Shao, J., 2010. Secure threshold multi authority attribute based encryption without a central authority. Information Sciences, 180(13), pp.2618-2632.
[56] Li, J., Huang, X., Li, J., Chen, X. and Xiang, Y., 2013. Securely outsourcing attribute-based encryption with checkability. IEEE Transactions on Parallel and Distributed Systems, 25(8), pp.2201-2210.
[57] Rouselakis, Y. and Waters, B., 2015, January. Efficient statically-secure large-universe multi-authority attribute-based encryption. In International Conference on Financial Cryptography and Data Security (pp. 315-332). Berlin, Heidelberg: Springer Berlin Heidelberg.
[58] Yu, S., Wang, C., Ren, K. and Lou, W., 2010, April. Attribute based data sharing with attribute revocation. In Proceedings of the 5th ACM symposium on information, computer and communications security (pp. 261-270).
[59] Wang, S., Zhou, J., Liu, J.K., Yu, J., Chen, J. and Xie, W., 2016. An efficient file hierarchy attribute-based encryption scheme in cloud computing. IEEE Transactions on Information Forensics and Security, 11(6), pp.1265-1277.
[60] Liu, Z., Cao, Z. and Wong, D.S., 2012. White-box traceable ciphertext-policy attribute-based encryption supporting any monotone access structures. IEEE Transactions on Information Forensics and Security, 8(1), pp.76-88.
[61] Hur, J. and Noh, D.K., 2010. Attribute-based access control with efficient revocation in data outsourcing systems. IEEE Transactions on Parallel and Distributed Systems, 22(7), pp.1214-1221.
[62] Qian, H., Li, J., Zhang, Y. and Han, J., 2015. Privacy-preserving personal health record using multi-authority attribute-based encryption with revocation. International Journal of Information Security, 14, pp.487-497.
[63] Han, J., Susilo, W., Mu, Y., Zhou, J. and Au, M.H.A., 2014. Improving privacy and security in decentralized ciphertext-policy attribute-based encryption. IEEE transactions on information forensics and security, 10(3), pp.665-678.
[64] Lewko, A. and Waters, B., 2012, August. New proof methods for attribute-based encryption: Achieving full security through selective techniques. In Annual Cryptology Conference (pp. 180-198). Berlin, Heidelberg: Springer Berlin Heidelberg.
[65] Garg, S., Gentry, C., Halevi, S., Sahai, A. and Waters, B., 2013. Attribute-based encryption for circuits from multilinear maps. In Advances in Cryptology–CRYPTO 2013: 33rd Annual Cryptology Conference, Santa Barbara, CA, USA, August 18-22, 2013. Proceedings, Part II (pp. 479-499). Springer Berlin Heidelberg.